\documentclass[10pt,twocolumn,conference]{IEEEtran}
\usepackage{times}
\usepackage{amsbsy}
\usepackage{amssymb}
\usepackage{fontenc}
\usepackage[usenames,dvipsnames]{color}
\usepackage{latexsym}
\usepackage{amsmath}
\usepackage[ansinew]{inputenc}
\usepackage[dvips ]{graphicx}
\input epsf
\usepackage{epic,eepic}
\usepackage{url}

\IEEEoverridecommandlockouts

\title{Multi-User Flexible Coordinated Beamforming using Lattice Reduction for Massive MIMO Systems}
\author{\textit{Keke Zu}$^1$, \textit{Bin Song}$^1$, \textit{Martin Haardt}$^1$,~\textit{and} \textit{Rodrigo C.\ de Lamare}$^2$\\
1~ Communications Research Laboratory, Ilmenau University of Technology \\
PO Box 100565, D-98684 Ilmenau, Germany \\
2~ Dept. of Electronics / CETUC, University of York / PUC-Rio\\
York YO10 5DD, U.K./ Marqu\^ede S.Vicente, 225 G\'avea, Rio de Janeiro, Brazil\\
Emails: zukeke@gmail.com, \{bin.song, martin.haardt\}@tu-ilmenau.de, rodrigo.delamare@york.ac.uk}

\begin{document}
\maketitle \thispagestyle{empty} \vspace*{-1.5em}

\begin{abstract}
The application of precoding algorithms in multi-user massive
multiple-input multiple-output (MU-Massive-MIMO) systems is restricted by the
dimensionality constraint that the number of transmit antennas has
to be greater than or equal to the total number of receive antennas.
In this paper, a lattice reduction (LR)-aided flexible coordinated
beamforming (LR-FlexCoBF) algorithm is proposed to overcome the
dimensionality constraint in overloaded MU-Massive-MIMO
systems. A random user selection scheme is integrated with the
proposed LR-FlexCoBF to extend its application to MU-Massive-MIMO
systems with arbitary overloading levels. Simulation results
show that significant improvements in terms of bit error rate (BER)
and sum-rate performances can be achieved by the proposed
LR-FlexCoBF precoding algorithm.
\end{abstract}

\section{Introduction}

The concept of multi-user massive
multiple-input multiple-out (MU-Massive-MIMO) systems has been developed recently in
\cite{Marzetta01}-\cite{Marzetta02} to bring huge improvements in
throughput and radiated energy efficiency with inexpensive, low
power components. This concept departs from the common understanding
of multi-antenna systems and considers a framework in which certain
nodes in the network are equipped with antenna arrays featuring a
large number of closely spaced radiating elements (several tens to a
few hundreds). By exploiting the potential of large spatial
dimensions, MU-Massive-MIMO can increase the capacity 10 times or
more and simultaneously improve the radiated energy-efficiency
\cite{Rusek}. However, the base stations (BSs) are sometimes assumed to
have an unlimited number of antennas \cite{Fernandes} or the number
of BS antennas per user is impractically large \cite{Hoydis}.

In this work, we focus on the design of downlink precoding
algorithms which have dimensionality constraints, also known as
overloaded MU-Massive-MIMO systems. As pointed in \cite{Spencer,BD},
the application of most precoding algorithms is restricted by the
dimensionality constraint that the number of transmit antennas has
to be greater than or equal to the total number of receive antennas.
Overloaded systems refer to those MIMO system scenarios in which the
dimensionality constraint is violated, i.e., the total number of
receive antennas exceeds the number of transmit antennas.

To overcome the dimensionality constraint,
receive antenna selection and eigenmode selection have been proposed
in \cite{Shen}. In both cases, however, the transmitter and the
receiver are not jointly optimized and some signaling techniques are
required to indicate the selected receive antennas or eigenmodes.
Alternatively, coordinated beamforming (CBF) techniques have been
developed to jointly update the transmit-receive beamforming vectors
\cite{Zhou}-\cite{Bin03}. However, a study of the convergence
behavior in terms of the number of iterations is not considered in
\cite{Zhou}, and the number of iterations is set manually. The
coordinated transmission strategy in \cite{Chae} only supports a
single data stream to each user. To support the transmission of
multiple data streams to each user and to reduce the number of
iterations, we have developed a flexible coordinated beamforming
(FlexCoBF) algorithm in \cite{Bin,Bin03}.

In order to implement FlexCoBF algorithm, a receive beamforming
matrix is employed at each user and is updated with the transmit
beamforming matrices iteratively. The receive beamforming matrices
may amplify the noise power at the receive side, resulting in
a reduced throughput and a degraded bit error rate (BER)
performance. To address this performance degradation, a lattice
reduction (LR) \cite{LLL} technique is employed and integrated with
the FlexCoBF to further improve the system performances. We term the ratio between
the total number of receive antennas and transmit antennas as
loading coefficient(LC). In this work, a random user selection scheme is
also developed for the proposed LR-FlexCoBF algorithm with arbitary
loading coefficients, which effectively extends the application of
FlexCoBF type precoding algorithms to arbitary overloading
scenarios. Moreover, current algorithms for MU-Massive-MIMO systems
assume that each user is equipped with a single antenna. However,
distributed users equipped with multiple antennas are likely to be
common since multiple antennas are already supported in LTE-Advanced
\cite{Lte} and in modern mobile devices \cite{iPhone}.

The main contributions of this work are summarized below:
\begin{enumerate}
 \item We study overloaded MU-Massive-MIMO systems. To the best of our knowledge,
 this is the first time this scenario is discussed for massive MIMO systems.

 \item A LR-aided FlexCoBF (LR-FlexCoBF) algorithm is proposed to overcome the
 dimensionality constraint problem in overloaded MU-Massive-MIMO systems.

 \item A user selection scheme is applied with the FlexCoBF algorithm to
 adjust the LC, resulting in a controlled sum-rate and BER performance.
\end{enumerate}

This paper is organized as follows. The system model and the LR-aided precoding algorithm are described in Section II and Section III, respectively. The proposed LR-FlexCoBF with random user selection algorithm is described in detail in Section IV . Simulation results and conclusions are presented in Section V and Section VI, respectively.

\section{System Model}

\begin{figure}[ht]
\begin{center}
\def\epsfsize#1#2{0.95\columnwidth}
\epsfbox{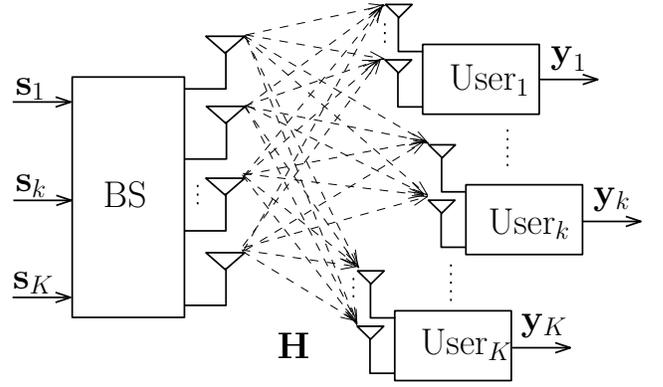} 
\caption{\footnotesize The MU-Massive-MIMO System Model} \label{MU_MIMO_System_Model}
\end{center}
\end{figure}

We consider an uncoded MU-Massive-MIMO broadcast system in a single cell environment as illustrated in Fig. \ref{MU_MIMO_System_Model},
equipped with $M_T$ transmit antennas at the base station (BS), $K$
users in the system each equipped with $M_k$ receive antennas, and
the total number of receive antennas is $M_R=\sum _{k=1}^{K}M_k$.
The combined transmit data streams are denoted as ${\boldsymbol
s}=[{\boldsymbol s^T_1},{\boldsymbol s^T_2},\cdots,{\boldsymbol
s^T_K}]^T\in\mathbb{C}^{r\times 1}$ with $\boldsymbol
s_k\in\mathbb{C}^{r_k\times 1}$, where $r$ is the total number of
transmit data streams and $r_k$ is the number of data streams for
user $k$. The combined channel matrix is denoted as ${\boldsymbol
H}=[{\boldsymbol H^T_1},{\boldsymbol H^T_2},\cdots,{\boldsymbol
H^T_K}]^T\in\mathbb{C}^{M_R\times M_T}$ and $\boldsymbol
H_k\in\mathbb{C}^{M_k\times M_T}$ is the $k$th user's channel
matrix. When channel state information (CSI) is available at the
transmit side, precoding techniques can be employed to pre-process
the transmit data and to reduce the multi-user interference (MUI)
\cite{Spencer}. Due to the large number of antennas at the BS, it is
very challenging to acquire CSI in frequency division
duplex (FDD) systems. For this reason, massive MIMO systems are
likely to operate in the time-division duplex (TDD) model, where the
reverse channel is used as an estimate of the forward channel
\cite{Marzetta01}-\cite{Marzetta02}. We assume a flat fading MIMO
channel and the received signal $\boldsymbol
y\in\mathbb{C}^{M_R\times 1}$ is given by
  \begin{align}
  \boldsymbol y =\boldsymbol H\boldsymbol F\boldsymbol s+ \boldsymbol n,
  \end{align}
where $\boldsymbol F=[\boldsymbol F_1 ~ \boldsymbol F_2
~ \ldots ~ \boldsymbol F_K]\in\mathbb{C}^{M_T\times r}$ is the
combined precoding matrix and ${\boldsymbol n=[\boldsymbol n^T_1 ~
\boldsymbol n^T_2 ~ \ldots ~ \boldsymbol
n^T_K]^T}\in\mathbb{C}^{M_R\times 1} $ is the combined Gaussian
noise with independent and identically distributed (i.i.d.) entries
of zero mean and variance $\sigma_n^2$.

\section{LR-aided Precoding Algorithm}

Lattice reduction (LR) can be considered as a mathematical theory to
find a basis with short, nearly orthogonal vectors for a given
integer lattice basis \cite{LLL}. Yao and Wornell \cite{Yao} first
applied the LR algorithm in conjunction with MIMO detection
techniques. In prior work \cite{Windpassinger}-\cite{Kek03} LR-aided MIMO
precoding algorithms have been devised and investigated. As shown
and studied in these works, the symbol error rate curves of
precoding and detection algorithms can approach the maximum
diversity order at the receive or transmit side.

Linear precoding can be interpreted as the problem of designing the
linear precoding matrix $\boldsymbol F$ to satisfy an optimization
criterion subject to a constraint associated with the transmit power
$E\{\|\beta \boldsymbol F\boldsymbol s\|\} \leq P_s$, where $P_s$ is
the total transmit power and the factor $\beta$ is chosen to scale
the transmitted power to $P_s$ \cite{Michael}. Based on the
optimization criterion, linear precoding algorithms can be
categorized into zero forcing (ZF) and minimum mean square error
(MMSE) based designs \cite{Michael}. Linear precoding algorithms are
attractive due to their simplicity. However, their transmit
diversity order is limited as compared to non-linear
precoding algorithms such as dirty paper coding (DPC) \cite{Costa},
Tomlinson-Harashima precoding (THP) \cite{Tomlinson, Harashima}, and
vector perturbation (VP) \cite{Hochwald, BDVP}. Although better
performance can be achieved by non-linear precoding algorithms,
their computational complexity is relatively high certain scenarios \cite{Kek03}.

With the aid of LR techniques, linear precoding techniques can
achieve the maximum diversity order while maintaining their
simplicity \cite{Windpassinger}. The most commonly used LR algorithm
is the LLL algorithm which was first proposed by Lenstra, Lenstra,
and L. Lov\'asz in \cite{LLL}. By using the LLL algorithm, only the
real-valued matrix can be processed which means the channel matrix has to be transformed into
equivalent double sized real-valued channel matrix. Thus extra unnecessary complexity could be introduced when the channel has large dimensions. In
order to reduce the computational complexity, the complex
LLL (CLLL) algorithm was proposed in \cite{CLLL}. The overall
complexity of CLLL algorithm is nearly half of the LLL algorithm
without sacrificing any performance. Therefore, we employ the CLLL
algorithm in this work. A complex lattice is a set of points
described by
\begin{align}
L(\boldsymbol H)=\{\boldsymbol H\boldsymbol x|x_l\in\mathbb{Z}+j\mathbb{Z}\},
\end{align}
where $\boldsymbol H=[\boldsymbol h_1,\boldsymbol h_2,\cdots,\boldsymbol h_{M_T}]$ contains the bases of the lattice $L(\boldsymbol H)$. The aim of the CLLL algorithm is to find a new basis $\boldsymbol {\tilde H}$ which is shorter and nearly orthogonal compared to the original matrix $\boldsymbol H$. Let us calculate the QR decomposition, $\boldsymbol H=\boldsymbol Q\boldsymbol R$, where $\boldsymbol Q$ is an orthogonal matrix and the upper-triangular matrix $\boldsymbol R$ is a rotated and reflected representation of $\boldsymbol H$. Thus, each column vector $\boldsymbol h_m$ of $\boldsymbol H$ is given by \cite{Wuebben}
\begin{align}
\boldsymbol h_m=\sum _{l=1}^{m}r_{l,m}\boldsymbol q_l,
\end{align}
where $\boldsymbol q_l$ is the $l$th column of $\boldsymbol Q$. If $|r_{1,m}|,..., |r_{m-1,m}|$ are close to zero, we can say that $\boldsymbol h_m$ is nearly orthogonal to the space spanned by $\boldsymbol h_1,...,\boldsymbol h_{m-1}$. Similarly, the QR decomposition of $\boldsymbol{\tilde H}$ is $ \boldsymbol{\tilde H}=\boldsymbol{\tilde Q}\boldsymbol{\tilde R}$. Then, the basis for $L(\boldsymbol H)$ is CLLL reduced if both of the following conditions are satisfied
\begin{align}
|\tilde r_{l,m}|\leq {1\over 2}|\tilde r_{l,l}|,\ 1\leq l<m\leq M_T,
\end{align}
\begin{align}
\delta|\tilde r_{m-1,m-1}|^2 \leq |\tilde r_{m,m}|^2+|\tilde r_{m-1,m}|^2, \ 2\leq m\leq M_T,
\end{align}
where $\delta \in ({1\over2},1]$ influences the quality of the
reduced basis and the computational complexity. We usually set
$\delta={3\over4}$ to achieve a trade-off between performance and
complexity \cite{LLL}.

We perform the LR transformation on the transpose of channel matrix
${\boldsymbol H}^T$ \cite{Windpassinger} as descripted by
\begin{align}
{\boldsymbol {\tilde H}}=\boldsymbol T\boldsymbol H ~{\rm and}~ \boldsymbol H=\boldsymbol T^{-1}{\boldsymbol {\tilde H}},
\end{align}
where $\boldsymbol T$ is a unimodular matrix with
$\rm{det}|\boldsymbol T|=1$ and all elements of $\boldsymbol T$ are
complex integers, i.e., $ t_{l,m} \in\mathbb{Z}+j\mathbb{Z}$. The
physical meaning of the constraint $\rm{det}|\boldsymbol T|=1$ is
that the channel energy is still the same after the LR
transformation. Following the LR transformation, we employ the
linear precoding constraint to get the precoding filter at the
transmit side and to process the data streams. The ZF precoding is implemented as
\begin{align}
\boldsymbol {\tilde F}_{\rm ZF}={\boldsymbol {\tilde H}}^H (\boldsymbol {\tilde H}{\boldsymbol {\tilde H}}^H)^{-1}={\boldsymbol H}^H (\boldsymbol H{\boldsymbol H}^H)^{-1}{\boldsymbol T}^{-1}.
\end{align}

\section{Proposed LR-aided FlexCoBF Algorithm with Random User Selection}

For linear or LR-aided precoding techniques to work, it is required
that $M_R\leq M_T$ \cite{Spencer}. When $M_R > M_T$, however,
precoding techniques cannot perform well simply because the
requirement of $M_R$ data streams is beyond the transmission ability
of the systems. For the $M_R > M_T$ case, we assume that the number
of actually transmitted data streams is $r$ and it should satisfy
$r\leq M_T$, which means that the maximum number of transmitted data
streams cannot exceed the number of transmit antennas $M_T$.

In \cite{Bin03}, we have developed an iterative coordinated method
named FlexCoBF to overcome the dimensionality constraint problem.
The receive beamforming matrix $\boldsymbol
W_k\in\mathbb{C}^{r\times M_k}$ is introduced at each user and
initialized with random matrices. Then, iterative computations are
employed to update the receive beamforming matrix $\boldsymbol W_k$ and the $k$th user's precoding
matrix $\boldsymbol F_k$ jointly to enforce the zero MUI constraint.
However, a random user selection scheme is developed in this work to
extend the application of FlexCoBF for MU-Massive-MIMO systems with
various loading coefficients. Considering the case when $M_T>K$, the
system can afford one data stream for each user plus extra data
streams for $M_T-K$ users equipped with multiple receive antennas.
Assume $j=M_T-K$, a random user selection scheme is first
implemented to select $j$ users from a total set of $K$ users. Then, the
remaining $l=K-j$ users are allocated one data stream by
implementing the iterative computations described above.
We define the quantity $\boldsymbol H_J=[{\boldsymbol H_{J_1}^T},{\boldsymbol H_{J_2}^T},\cdots,{\boldsymbol
H_{J_j}^T}]^T$ is the combined channel matrix
of the selected users, and $\boldsymbol H_L=[{\boldsymbol H_{L_1}^T},{\boldsymbol H_{L_2}^T},\cdots,{\boldsymbol
H_{L_l}^T}]^T$ is the combined channel
matrix of the remaining users.
Finally, the equivalent channel matrix ${\boldsymbol
H_e}\in\mathbb{C}^{r\times M_T}$ is obtained as
\begin{align}
\boldsymbol H_e=\begin{bmatrix} \boldsymbol I_1 & 0 & 0 & 0 & \ldots& 0\\ \vdots & \ddots& \vdots & \vdots& \vdots& \vdots
\\ 0 & 0 & \boldsymbol I_j & 0 & \ldots& 0 \\ 0 & 0 & 0 & \boldsymbol W_1^H & \ldots& 0 \\\vdots & \vdots& \vdots & \vdots& \ddots& \vdots\\ 0 & 0 & 0 &0 & \ldots & \boldsymbol W_l^H \end{bmatrix}
\begin{bmatrix} \boldsymbol H_{J_1} \\ \vdots \\ \boldsymbol H_{J_j} \\ \boldsymbol H_{L_1} \\ \vdots\\\\ \boldsymbol H_{L_l}  \end {bmatrix}.
\end{align}

Since the standard lattice reduction algorithm is also restricted by
the dimensionality constraint \cite{CLLL}, we apply the LR
transformation on the equivalent channel matrix ${\boldsymbol H_e}$
rather than the channel matrix ${\boldsymbol H}$ when $M_R > M_T$,
i.e.,
\begin{align}
\boldsymbol{\tilde H}_e=\boldsymbol T_e\boldsymbol H_e,
\end{align}
where the quantity $\boldsymbol{\tilde H}_e\in\mathbb{C}^{r\times
M_T}$ is the LR transformed matrix and the transformation matrix
$\boldsymbol T_e\in\mathbb{C}^{r\times r}$ is unimodular ($|{\rm
det} (\boldsymbol T_e)|=1$) and all elements of $\boldsymbol T_e$
are complex integers. The LR transformed matrix $\boldsymbol {\tilde
H}_e$ is a basis of ${\boldsymbol H_e}$ in the lattice space.
Compared to the original equivalent channel matrix $\boldsymbol
H_e$, the LR transformed channel matrix $\boldsymbol {\tilde H}_e$
is closer to orthogonality which can be measured by the orthogonality
defect factor defined in \cite{ODF}. Therefore, improved BER and sum-rate
performances can be achieved by the proposed LR-FlexCoBF algorithm.

Assuming that the variable $p$ represents the iteration index, the proposed LR-FlexCoBF algorithm is performed in the following seven steps:
\begin{enumerate}
 \item Implement the random user selection scheme to select $J$ users from $K$, and get $\boldsymbol H_J$ and $\boldsymbol H_L$, respectively.
  \item Initialize the iteration index $p$ to zero and the combined receive beamforming matrix $\boldsymbol W^{(0)}={\rm diag}\{{\boldsymbol W_1^{(0)}}^H, {\boldsymbol W_2^{(0)}}^H, \cdots, {\boldsymbol W_l^{(0)}}^H\}$ to random  matrices. Set the constant $\epsilon$ as the threshold to iteratively enforce the zero MUI constraint for each receiver.
   \item Set $p=p+1$ and compute the equivalent channel matrix $\boldsymbol H_{L_e}^{(p)}$ as
          \begin{align}
                    \boldsymbol H_{L_e}^{(p)}&=\begin {bmatrix} {\boldsymbol W_1^{(p-1)}}^H\boldsymbol H_{L_1}\\ {\boldsymbol W_2^{(p-1)}}^H\boldsymbol H_{L_2}\\ \vdots  \\ {\boldsymbol W_K^{(p-1)}}^H\boldsymbol H_{L_l}\end {bmatrix}.\nonumber
                \end{align}

     \item Apply the ZF constraint based precoding design to the obtained equivalent channel matrix $\boldsymbol H_{L_e}^{(p)}$ to obtain the transmit beamforming matrices
     $\boldsymbol  F_k^{(p)}$ for all single data stream users ($k=1,\ldots,l$).

   \item For the $p$th iteration, update $\boldsymbol W^{(p)}$
   jointly with the newly obtained precoding matrix $\boldsymbol  F_{L_e}^{(p)}$ as
             \begin{align}
                    {\boldsymbol W^{(p)}}&=\boldsymbol H\boldsymbol  F_{L_e}^{(p)}, \nonumber
         \end{align}
where the quantity $\boldsymbol H$ is the combined channel matrix defined in Section II.

    \item Track the alterations of the residual MUI after the linear precoding as
               \begin{align}
                    {\rm MUI} (\boldsymbol H_{L_e}^{(p+1)}{\boldsymbol F_{L_e}^{(p)}})=\|{\rm off}(\boldsymbol H_{L_e}^{(p+1)}{\boldsymbol F_{L_e}^{(p)}})\|_F^2, \nonumber
         \end{align}
where the operation ${\rm off}(\boldsymbol B)$ denotes the selection
of the off-diagonal elements of the matrix $\boldsymbol B$. If the
residual MUI is above the threshold $\epsilon$, go back to step 2.
Otherwise, convergence is achieved and the iterative procedure
stops. The final equivalent channel matrix $\boldsymbol H_e$ and the
receive beaforming matrix $\boldsymbol W_e^{(p)}$ is respectively
obtained as $\boldsymbol H_e=[{\boldsymbol H_J^T}, {\boldsymbol
H_{L_e}^{(p)}}^T]^T$ and $\boldsymbol W_e^{(p)}={\rm diag}\{{\boldsymbol
I_J},{\boldsymbol W^{(p)}}\}$.
    \item Perform the LR transformation on the obtained equivalent
    channel matrix $\boldsymbol H_e$ to get $\boldsymbol {\tilde H}_e$
    and the transformation matrix $\boldsymbol T_e$. Apply the
    transformation matrix $\boldsymbol T_e$ to the obtained
    combined precoding matrix
    ${\boldsymbol F_e^{(p)}}=[\boldsymbol F_1^{(p)}, \boldsymbol F_2^{(p)},\ldots,\boldsymbol F_K^{(p)}]$ ,
    that is, $\boldsymbol {\tilde F}_e^{(p)}={\boldsymbol F_e^{(p)}}{\boldsymbol T_e}^{-1}$.
 \end{enumerate}
Finally, the received signal for the $M_R > M_T$ case can be expressed as
  \begin{align}
  \boldsymbol y =\boldsymbol W_e^{(p)}\boldsymbol H\boldsymbol {\tilde F}_e^{(p)}\boldsymbol s+ \boldsymbol n=\boldsymbol W_e^{(p)}\boldsymbol H\boldsymbol F_e^{(p)}\boldsymbol z+ \boldsymbol n,
  \end{align}
where the quantity $\boldsymbol z={\boldsymbol T_e}^{-1}\boldsymbol s$. In order to get the estimation of $\boldsymbol z$, a proper shifting and scaling work is needed to transform the signal back from the Lattice space \cite{Kek04}. Then, the estimated transmit signal $\boldsymbol s$ is obtained as $\boldsymbol {\hat s}=\boldsymbol T \boldsymbol {\hat z}$ with $\boldsymbol {\hat z}$ is the estimation of $\boldsymbol z$.


\section{Simulation Results}

We consider an overloaded MU-Massive-MIMO system in this section.
The quantity $E_b/N_0$ is defined as $E_b/N_0={M_R P_s\over
M_TN\sigma_n^2}$ with $N$ being the number of information bits
transmitted per channel symbol, and the loading coefficient ($L_C$) is
defined as $L_C={M_R\over M_T}$. An uncoded QPSK modulation scheme is
employed in the simulations. The threshold $\epsilon$ is set to
$10^{-5}$, and the maximum number of iterations is restricted to
$20$. The channel matrix ${\boldsymbol H}$ is assumed to be a
complex i.i.d. Gaussian matrix with zero mean and unit variance.

\begin{figure}[htp]
\begin{center}
\def\epsfsize#1#2{0.95\columnwidth}
\epsfbox{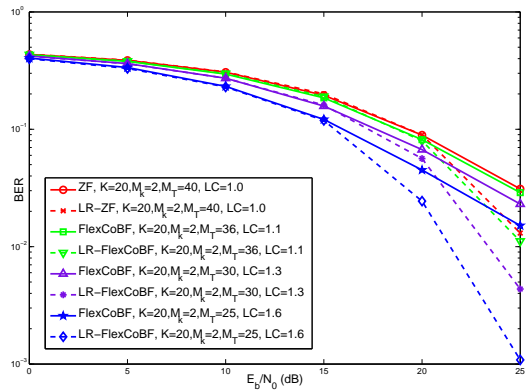} \vspace{-0.8em} \caption{\footnotesize BER
performance with QPSK, $K=20$, $M_k=2$} \label{40_BER}
\end{center}
\end{figure}

\begin{figure}[htp]
\begin{center}
\def\epsfsize#1#2{0.95\columnwidth}
\epsfbox{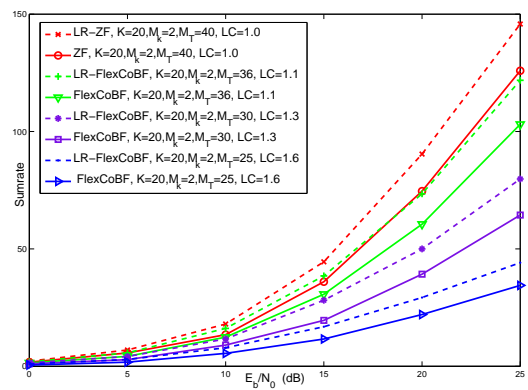} \vspace{-0.8em} \caption{\footnotesize Sum-rate
performance, $K=20$, $M_k=2$} \label{40_Capacity}
\end{center}
\end{figure}

The case of $K=20$ users each equipped with $M_k=2$ receive antennas is first considered in Fig.
\ref{40_BER} and Fig. \ref{40_Capacity}. When the loading
coefficient $L_C=1.0$, FlexCoBF corresponds to conventioanl ZF
precoding. It is shown in Fig. \ref{40_BER} that the BER performance
becomes better with the increase of $L_C$.
This is because that the inter-stream interference is reduced with the
decrease of the equivalent channel matrix dimension. The proposed
LR-FlexCoBF has a better performance as compared to the conventional
FlexCoBF especially for the heavily overloaded case at high SNRs.
The sum-rate performance, which is illustrated in Fig. \ref{40_Capacity},
is inversely proportional to $L_C$ because more data streams are supported. For example,
$40$ data streams are supported with $L_C=1.0$ while there are only
$25$ with $L_C=1.6$. It worth noting that a much better sum-rate
performance is achieved by the proposed LR-FlexCoBF algorithms.

\begin{figure}[htp]
\begin{center}
\def\epsfsize#1#2{0.95\columnwidth}
\epsfbox{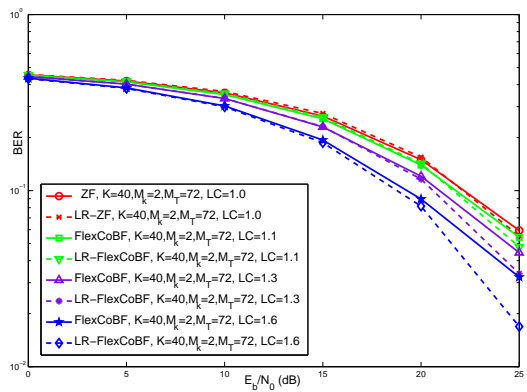} \vspace{-0.8em} \caption{\footnotesize BER
performance with QPSK, $K=40$, $M_k=2$} \label{80_BER}
\end{center}
\end{figure}

\begin{figure}[htp]
\begin{center}
\def\epsfsize#1#2{0.95\columnwidth}
\epsfbox{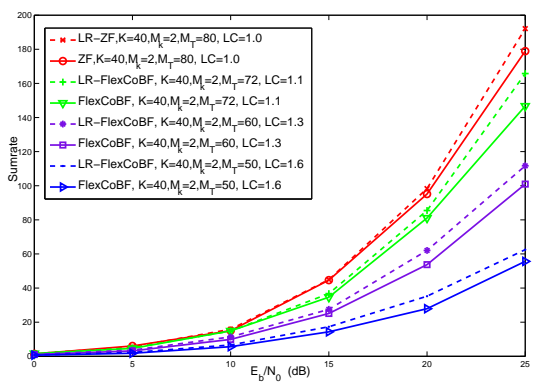} \vspace{-0.8em} \caption{\footnotesize Sum-rate
performance,  $K=40$, $M_k=2$} \label{80_Capacity}
\end{center}
\end{figure}

The BER and sum-rate performance of $K=40$ with $M_k=2$ is illustrated in Fig.
\ref{80_BER} and Fig. \ref{80_Capacity}, respectively. Similar
performances are observed as displayed in Fig. \ref{40_BER} and Fig.
\ref{40_Capacity}. It should be noted that for the heavily
overloaded case with same $L_C$, the BER
performance degraded faster than the curve in Fig. \ref{40_BER}.

\section{Conclusion}

A LR-FlexCoBF algorithm has been proposed to support data transmission
in overloaded MU-Massive-MIMO system. By employing a random user
selection scheme in conjuction with the proposed LR-FlexCoBF algorithm, MU-Massive-MIMO systems with varity loading
coefficients are studied. The proposed LR-FlexCoBF precoding
algorithm can achieve a higher diversity order and a higher spatial
multiplexing gain as compared to the standard FlexCoBF and other
existing techniques.

\end{document}